\documentclass[aps,prd,onecolumn,amsfonts,amssymb,amsmath,showpacs,floatfix,nofootinbib,groupedaddress,superscriptaddress,citesort]{revtex4}
\usepackage{mathrsfs}
\usepackage{amsfonts}
\usepackage{amstext}
\usepackage{amsmath}
\usepackage{amssymb}
\usepackage{float}
\usepackage[colorlinks, citecolor=blue]{hyperref}
\usepackage[usenames]{xcolor}
\usepackage[dvips]{graphicx}
\def\qed{\leavevmode\unskip\penalty9999 \hbox{}\nobreak\hfill
     \quad\hbox{\leavevmode  \hbox to.77778em{%
              \hfil\vrule   \vbox to.675em%
               {\hrule width.6em\vfil\hrule}\vrule\hfil}}
     \par\vskip3pt}

\def\ra{\rangle}
\def\la{\langle}

\def\no{\nonumber}
\def\bea{\begin{eqnarray}}
\def\eea{\end{eqnarray}}
\def\be{\begin{equation}}
\def\ee{\end{equation}}

\begin{document}
\title{Uncertainty relation in Schwarzschild spacetime}

\author{Jun Feng}
\email{j.feng1@uq.edu.au}
\affiliation{Department of Applied Physics, Xi'an Jiaotong University, Xi'an 710049, PR China}
\affiliation{School of Mathematics and Physics, The University of Queensland, Brisbane, QLD 4072, Australia}
\author{Yao-Zhong Zhang}
\affiliation{School of Mathematics and Physics, The University of Queensland, Brisbane, QLD 4072, Australia}
\author{Mark D. Gould}
\affiliation{School of Mathematics and Physics, The University of Queensland, Brisbane, QLD 4072, Australia}
\author{Heng Fan}
\affiliation{Institute of Physics, Chinese Academy of Sciences, Beijing 100190, PR China}

\begin{abstract}

We explore the entropic uncertainty relation in the curved background outside a Schwarzschild black hole, and find that Hawking radiation introduces a nontrivial modification on the uncertainty bound for particular observer, therefore it could be witnessed by proper uncertainty game experimentally. We first investigate an uncertainty game between a free falling observer and his static partner holding a quantum memory initially entangled with the quantum system to be measured. Due to the information loss from Hawking decoherence, we find an inevitable increase of the uncertainty on the outcome of measurements in the view of static observer, which is dependent on the mass of the black hole, the distance of observer from event horizon, and the mode frequency of quantum memory. To illustrate the generality of this paradigm, we relate the entropic uncertainty bound with other uncertainty probe, e.g., time-energy uncertainty. In an alternative game between two static players, we show that quantum information of qubit can be transferred to quantum memory through a bath of fluctuating quantum fields outside the black hole. For a particular choice of initial state, we show that the Hawking decoherence cannot counteract entanglement generation after the dynamical evolution of system, which triggers an effectively reduced uncertainty bound that violates the intrinsic limit $-\log_2c$. Numerically estimation for a proper choice of initial state shows that our result is comparable with possible real experiments. Finally, a discussion on the black hole firewall paradox in the context of entropic uncertainty relation is given. 

\end{abstract}
\maketitle

\section{Introduction}
\label{1}

One of the most remarkable features of quantum physics is Heisenberg's uncertainty principle, which bounds the ability of an observer holding \emph{classical} information to precisely predict the outcomes of incompatible measurements on a quantum particle. For incompatible measurements $Q$ and $R$, the uncertainty relation can be recast by entropic measures as \cite{H1,H2,H3} $H(Q)+H(R)\geqslant-\log_2c$, with $H$ the Shannon entropy for the probability distribution of the measurement outcomes. The overlap between observables $Q$ and $R$ is a constant $c=\mbox{max}_{i,j}|\la a_i| b_j\ra|^2$ with $|a_i\ra$ and $|b_j\ra$ the corresponding eigenvectors, thus providing a state-independent lower bound indicating an intrinsic uncertainty in any quantum systems. However, once the observer has access to previously determined \emph{quantum} information of measured system $A$, stored in a quantum memory $B$ which is entangled with the system, the uncertainty bound can be dramatically violated \cite{EUR1}. Quantitatively, the quantum-memory-assisted entropic uncertainty relation (EUR) can be expressed as
\be
S(Q|B)+S(R|B)\geqslant-\log_2c+S(A|B)
\label{eur2}
\ee
where the quantum conditional von Neumann entropy $S(A|B)=S(\rho_{AB})-S(\rho_{B})$ could be negative \cite{EUR2} and has significant operational meaning \cite{EUR3}. In the extreme case where $A$ and $B$ are maximally entangled, the right-hand side (r.h.s.) is vanishing, enabling us to predict the outcomes precisely. Otherwise, if $A$ and $B$ are not entangled, the constant bound $-\log_2c$ is recovered. In this meaning, as been experimentally verified \cite{EUR4}, the negative conditional entropy is a sufficient condition for entanglement \cite{EUR5}, therefore the quantum-memory-assisted EUR (\ref{eur2}) can be used to efficiently witness quantum entanglement.

Falling in a relativistic realm, we can expect \cite{FENG} that the quantum-memory-assisted form of uncertainty relation would be inevitable, since quantum correlations in relativistic context is highly observer-dependent (see review \cite{RQI1} and references therein). With the appearance of causal horizons, quantum entanglement exhibits a degradation phenomenon for a noninertial observer due to the Unruh effect in flat space \cite{RQI2,RQI5}, or for a static observer located at a fixed distance from the event horizon of black hole \cite{RQI3,RQI4}. The behavior of entanglement in a dynamical spacetime has also been investigated \cite{RQI6,RQI7,RQI8,RQI9}. This environment decoherence should certainly modify the above quantum-memory-assisted uncertainty bound \cite{APP}. For instance, an uncertainty game between two players Alice and Bob in noninertial frame has been explored in \cite{FENG}, which shows an increasing uncertainty in bipartite system $\rho_{AB}$ of free Dirac field, and a periodic evolution of uncertainty for localized quantum system restricted in cavities, providing an efficient relativistic entanglement witness that could be detected experimentally \cite{EUR4}.

In this Letter, we explore the EUR (\ref{eur2}) in the curved background outside a Schwarzschild black hole, and show that Hawking radiation introduce a nontrivial modification on the uncertainty bound for particular observer, therefore could be witnessed in a proper uncertainty game in principle \cite{EUR4}. In an uncertainty game between a free falling player and a static player, we show that the entanglement between quantum memory and the measured quantum system should be degraded by Hawking radiation, therefore resulting an inevitably increase of the uncertainty on the outcome of measurements. We give the quantum-memory-assisted uncertainty bound explicitly, which is dependent on three physical parameters, i.e., the mass of the black hole, the distance of observer from event horizon, and the mode frequency of quantum memory. We show that the most significant modification on uncertainty bound happens in the vicinity of event horizon, and becomes ignorable as both observers leave away from the black hole. Moreover, one can also utilize a quantum memory with high enough frequency to make a precisely prediction on the outcomes of incompatible measurements. We illustrate the generality of our results by comparing them with other uncertainty probe. In particular, an explicit link to Aharonov-Anandan time-energy uncertainty \cite{AAI1} is given.

In an alternative game between two static players outside a Schwarzschild black hole, we show that the uncertainty caused by Hawking decoherence can be counteracted and even overcome. By the interaction with a bath of fluctuating quantum fields outside the black hole, entanglement could be generated between mutually independent system after a quantum dynamical evolution \cite{YU1}. In this meaning, the quantum information of qubit can be transferred and stored in quantum memory which triggers a negative conditional von Neumann entropy $S(A|B)$, and enables a \emph{violation} of uncertainty bound $-\log_2c$ for particular player. Beside a similar dependence on the physical parameters in the first scenario, we show that the uncertainty bound of final equilibrium system is also sensitive with the choice of initial bipartite state of quantum memory and the qubit to be measured. By choosing a particular Bell-diagonal state as initial state, we show that the uncertainty bound can be effectively reduced and violate the intrinsic bound $-\log_2c$ after quantum dynamical evolution. Numerically estimation shows that our results are comparable with possible real experiments.

\section{Uncertainty game between free falling player and his static partner}
\label{2}

\subsection{Vacuum states in Schwarzschild spacetime}
To address an uncertainty game in the background of a black hole, we first recall the definition of vacuum states in the curved space. We consider a Schwarzschild black hole described by  
\be
ds^2=-\bigg(1-\frac{2M}{r}\bigg)dt^2+\bigg(1-\frac{2M}{r}\bigg)^{-1}dr^2+r^2d\mathbf{\Omega}^2
\label{bh1}
\ee 
where $M$ is the mass of black hole and $d\mathbf{\Omega}^2$ is the line element in the unit sphere. With the presence of event horizon at $R_H=1/2M$, the observer outside the black hole has no access to the information about those particle states inside the horizon. In terms of thermo-field dynamics \cite{BOG1}, this information-loss results a thermal character of the vacuum perceived by stationary observer, related to a spectrum with Hawking temperature $T_H=\kappa/2\pi$ where surface gravity is $\kappa=\frac{1}{4M}$. 

Consider a massless Dirac field which satisfies  $[i\gamma^\mu(\partial_\mu-\Gamma_\mu)]\psi=0$, where $\Gamma_\mu$ is spin connection. To specify its vacuum structure for different observer, we introduce the Kruskal coordinates
\be
U=-4M\exp[-\kappa(t-r^*)]\ ,\  V=4M\exp[\kappa(t+r^*)]
\ee
where $r^*=r+2M\ln|1-r/2M|$ is tortoise coordinate. In terms of the new coordinates, the maximal analytic extension of (\ref{bh1}) is
\be
ds^2=-\frac{1}{2\kappa r}e^{-2\kappa r}dUdV+r^2(d\theta^2+\sin^2\theta d\phi^2)
\label{bh2}
\ee
which has a near-horizon structure approximated by Rindler manifold. This enables us to analyze the vacuum structure of quantum field similar as for accelerated observer in flat space \cite{RQI4,BOG2}. For a free infalling observer (Alice) with proper timelike vector $\partial_{\hat{t}}\propto(\partial_U+\partial_V)$, the Hartle-Hawking vacuum $|0_H\ra$ is analogous to Minkowski vacuum. For an observer (Bob) with static timelike Killing vector $\partial_{t}\propto(U\partial_U-V\partial_V)$, the Boulware vacuum $|0\ra_I$ corresponding to positive frequencies associated to $\partial_{t}$ is analogous to Rindler vacuum in flat space. Similarly, another vacuum $|0\ra_{II}$ related with $-\partial_t$ could be find.

Nevertheless, one should be very careful on the validity of above analogy, since it can only work in the vicinity of the black hole. For Bob resisting in a position $r_0$ close enough to the event horizon, his proper acceleration is $a=\kappa/\sqrt{1-2M/r_0}$. This provides a constraint that Rindler approximation could only make sense if $r_0/R_H-1\ll1$ \cite{BOG3}. Taking into account that $|0_H\ra\equiv\bigotimes_i|0_{\omega_i}\ra_H$ and its first excitation $|1_H\ra\equiv\bigotimes_i|1_{\omega_i}\ra_H$, we can express Hartle-Hawking vacuum and its excitation in Boulware basis as \cite{RQI5,BOG3}
\bea
|0_{\omega_i}\ra_H&=&\big[1+\exp\big(-\Omega\sqrt{1-1/R_0}\;\big)\big]^{-\frac{1}{2}}|0_{\omega_i}\ra_I|0_{\omega_i}\ra_{II}\no\\
&&+\big[1+\exp\big(\Omega\sqrt{1-1/R_0}\;\big)\big]^{-\frac{1}{2}}|1_{\omega_i}\ra_I|1_{\omega_i}\ra_{II},\no\\
|1_{\omega_i}\ra_H&=&|1_{\omega_i}\ra_I|0_{\omega_i}\ra_{II}
\label{bh3}
\eea
where $R_0=r_0/R_H=r_0/2M$ and $\Omega=2\pi\omega/\kappa=8\pi\omega M$ is the mode frequency measured by Bob in terms of the surface gravity. It should be emphasized that the parameterization in (\ref{bh3}) allows us to analyze the uncertainty bound as a function of the distance of Bob to the event horizon, besides its dependence simply on Hawking temperature \cite{RQI4,BOG2}.

\subsection{Uncertainty bound in Schwarzschild spacetime}
We now turn to an uncertainty game between Alice and Bob \cite{EUR1}, where Bob sends Alice a qubit $A$, initially entangled with his quantum memory $B$. After Alice measuring either $Q$ or $R$ and broadcasting her
measurement choice, Bob needs to minimize his uncertainty about Alice¡¯s measurement outcome. To investigate the influence of Hawking effect on this game, we consider the bipartite system of Alice and Bob, initially sharing a Bell-diagonal state in a free falling basis
\be
\rho_{AB}=\frac{1}{4}(\mathbf{1}^A\otimes\mathbf{1}^B+\sum_{i=1}^{3}c_i\sigma_i^A\otimes\sigma_i^B)
\label{bell}
\ee
where $\sigma_i$ are Pauli matrix and the coefficients $0\leqslant|c_i|\leqslant1$. These states are the convex combination of four Bell states and reduce to maximally entangled states (Bell-basis) if $|c_1|=|c_2|=|c_3|=1$.The states (\ref{bell}) are the generalization of many important quantum states in quantum information theory, e.g., Werner state \cite{werner}, and have important experimentally application in testing EUR \cite{EUR2}. Following \cite{RQI3}, we assume that Alice has a detector which only detects mode with frequency $k$ and Bob has a detector sensitive to mode $\omega$. After their coincidence, while Alice remains free falling into the black hole,  Bob falls toward the black hole and then locates at a fixed distance $r_0$ outside the event horizon. Therefore, the states corresponding to mode $\omega$ must be specified in Boulware basis. Since the static observer cannot access the modes beyond the horizon, the lost information reduces the entanglement between $A$ and $B$, therefore modifies the uncertainty bound.

In particular, after transforming Bob's states according to (\ref{bh3}) and tracing over all degrees in region II, we obtain a reduced density matrix for bipartite system involving free falling observer Alice and static observer Bob with quantum memory
\bea
\rho_{AI}&=&\frac{1}{4}\big[1+e^{-\Omega\sqrt{1-1/R_0}}\big]^{-1}\bigg\{c_3\sigma_3^A\otimes\sigma_{I,3}^B+\mathbf{1}^A\otimes|0\ra_I\la0|+\frac{3+e^{\Omega\sqrt{1-1/R_0}}+2e^{-\Omega\sqrt{1-1/R_0}}}{1+e^{\Omega\sqrt{1-1/R_0}}}\mathbf{1}^A\otimes|1\ra_I\la1|\no\\
&&+\sum_{k=1}^2c_k\big[1+e^{-\Omega\sqrt{1-1/R_0}}\big]^{1/2}\sigma_k^A\otimes\sigma_{I,k}^B\bigg\}
\label{bacc}
\eea
where $\sigma_{I,1}^B=|0\ra_I\la1|+|1\ra_I\la0|$, $\sigma_{I,2}^B=-i|0\ra_I\la1|+i|1\ra_I\la0|$ and $\sigma_{I,3}^B=|0\ra_I\la0|-|1\ra_I\la1|$ are Pauli matrix constructed in Boulware basis.

\begin{figure}[hbtp]
\begin{center}
\includegraphics[width=.45\textwidth]{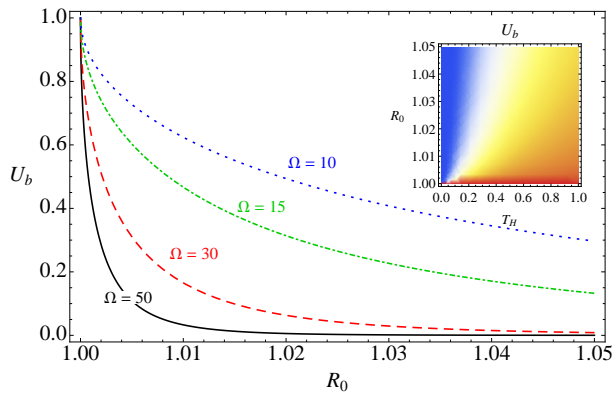}
\caption{The uncertainty bound $U_b$ is a function of $T_H$ and $R_0$. The maximally entangled state with $c_1=-c_2=c_3=1$ has been chosen, and the relative distance of Bob to event horizon $r_0/R_H\lesssim0.05R_H$ is assumed thus Rindler approximation can be hold. In the inset we plot the dependence of $U_b$ to Hawking temperature, where red color denotes higher $U_b$. In particular, at $R_0=1$ we have $U_b=1$. }
\label{ubh2}
\end{center}
\end{figure}

To specify the uncertainty bound on the r.h.s. of (\ref{eur2}), we need to calculate the conditional von Neumann entropy of (\ref{bacc}) that is $S(A|I)=S(\rho_{AI})-S(\rho_I)$. The reduced density matrix $\rho_{AI}$ has an X-type structure with eigenvalues  
\bea
\lambda_1=\frac{1}{4(1+e^{-\Omega\sqrt{1-1/R_0}})}[1+e^{-\Omega\sqrt{1-1/R_0}}+c_3+\xi_-]&,&\no\\
\lambda_2=\frac{1}{4(1+e^{-\Omega\sqrt{1-1/R_0}})}[1+e^{-\Omega\sqrt{1-1/R_0}}+c_3-\xi_-]&,&\no\\
\lambda_3=\frac{1}{4(1+e^{-\Omega\sqrt{1-1/R_0}})}[1+e^{-\Omega\sqrt{1-1/R_0}}-c_3+\xi_+]&,&\no\\
\lambda_4=\frac{1}{4(1+e^{-\Omega\sqrt{1-1/R_0}})}[1+e^{-\Omega\sqrt{1-1/R_0}}-c_3-\xi_+]&,&\no
\eea
where $\xi_\pm=\sqrt{(c_1\pm c_2)^2(1+e^{-\Omega\sqrt{1-1/R_0}})+e^{-2\Omega\sqrt{1-1/R_0}}}$. The von Neumann entropy of (\ref{bacc}) is then given by $S(\rho_{AI})=-\sum_{i=1}^4\lambda_i\log_2\lambda_i$. Further, for the quantum memory positioned at $r_0$ with $\rho_I=\mbox{Tr}_A\rho_{AI}$, the associated entropy is
\be
S(\rho_I)=H_{\tiny\mbox{bin}}\bigg(\frac{1}{2(1+e^{-\Omega\sqrt{1-1/R_0}})}\bigg)
\ee
where the binary entropy $H_{\tiny\mbox{bin}}(p)\equiv -p\log_2p-(1-p)\log_2(1-p)$ is defined. On the other hand, once Alice decide what measurement she would make, the complementarity $c$ of the observables $Q$ and $R$ is a fixed constant, provides a intrinsic part of uncertainty in (\ref{eur2}). In our game, we assume that the qubit $A$ of Alice is measured by one of the Pauli operators. For any two observables $\sigma_j$ and $\sigma_k$ ($j\neq k=1,2,3$), the complementarity is always $\frac{1}{2}$. Denoting the r.h.s. of (\ref{eur2}) as $U_b$, we finally have
\be
U_b=1-\sum_{i=1}^4\lambda_{i}\log_2\lambda_{i}-H_{\tiny\mbox{bin}}\bigg(\frac{1}{2(1+e^{-\Omega\sqrt{1-1/R_0}})}\bigg)
\label{rhs1}
\ee
which is clearly dependent on the mass of the black hole $M$, the distance of Bob from event horizon $R_0$, and the mode frequency $\omega$ of quantum memory. We illustrate the uncertainty bound (\ref{rhs1}) in Fig.\ref{ubh2}, where a particular choice of maximally entangled Bell state with $c_1=-c_2=c_3=1$ has been made. 

Firstly, we observe that for static observer Bob, the uncertainty bound $U_b$ is sensitive with the distance of his position to the event horizon. For $r_0=R_H$, quantum memory is disentangled with the system, we have $U_b=1$ agreeing with the intrinsic part of (\ref{eur2}). As both observers leave away from the horizon, the modification on uncertainty bound becomes ignorable, consequently, uncertainty game in an universes containing event horizons are not jeopardized. We also note that it is enough to investigate the modification in the vicinity of event horizon, e.g., $r_0/R_H\lesssim0.05R_H$, where Rindler approximation  works and all characterful behavior of $U_b$ would appear. 

Even in the proximities of a Schwarzschild black hole, we show that uncertainty bound of measurements could still be drastically reduced with growing $\Omega$. Since $\Omega=2\pi\omega/\kappa=\omega/T_H$, we conclude that outside a black hole with fixed mass (i.e., fixed Hawking temperature), static observer can predict precisely the outcomes of incompatible measurements on a quantum particle by choosing field modes with enough high frequency $\omega$ to hold. Moreover, since the significant change on $U_b$ observed by static observer matches the accuracy of current technology \cite{EUR2}, entropic uncertainty bound (\ref{rhs1}) indeed provides an efficiently witness of Hawking radiation from the event horizon. On the other hand, as depicted in the inset of Fig.\ref{ubh2}, where red color denotes higher $U_b$, we find the uncertainty bound $U_b$ always increase with higher Hawking temperature. This just means that Hawking radiation is essentially plays a role of environment decoherence which increases the uncertainty in our prediction of measurement outcomes \cite{RQI3,RQI4,BOG3}.

\subsection{Relationship between entropic uncertainty and Aharonov-Anandan time-energy uncertainty}

We would like to illustrate the generality of our results by comparing them with another uncertainty measure, i.e., Aharonov-Anandan (AA) time-energy uncertainty \cite{AAI1}, recently proposed to probe Hawking effect simulated by graphene morphologies in lab \cite{AAI2}. In particular, during the time evolution of a nonstationary quantum state $|\phi(t)\ra$, there is a nonzero energy uncertainty $\Delta E^2(t)=\la\phi(t)|H^2|\phi(t)\ra-(\la\phi(t)|H|\phi(t)\ra)^2$. Specify to a evolution from initial to final states by Bogoliubov transformation (\ref{bh3}) $|\Psi(\theta)\ra=J^{-1}(\theta)|\psi(t)\ra$, we then have the corresponding AA energy uncertainty $\Delta_{AA}\equiv\Delta E^2(\theta)$ charactering Hawking effect. Here $J(\theta)$ is generator of transformation (\ref{bh3}) with $\theta\equiv(\Omega,R_0,t)$ \cite{RQI5} and can be expressed as
\be
J(\theta)=\exp[\arctan(e^{-\Omega\sqrt{1-1/R_0}})(\hat{a}^{I\dag} \hat{c}^{II\dag}+\hat{a}^{I} \hat{c}^{II})]
\ee
where $\hat{a}^{\Sigma}$ and $\hat{c}^\Sigma$ ($\Sigma=\{I,II\}$) are fermionic particle and antiparticle operators annihilating Boulware vacuum $|0\ra_I$ and $|0\ra_{II}$. After some standard but tedious calculation \cite{BOG5}, we obtain
\be
\Delta_{AA}(\Omega,R_0)=\sqrt2\omega\big(2+e^{-\Omega\sqrt{1-1/R_0}}+e^{\Omega\sqrt{1-1/R_0}}\big)^{-\frac{1}{2}}
\label{AA1}
\ee
which is a product of coefficients in transformation (\ref{bh3}) \cite{AAI2} and have similar dependence on $\Omega$ and $R_0$ as in Fig. \ref{ubh2}. Therefore we could use this AA energy uncertainty as building block to reexpress our entropic uncertainty bound (\ref{rhs1}). For instance, for the maximally entangled state with $c_1=-c_2=c_3=1$, we have
\be
U_b=\log_2\frac{8}{3}+\frac{1}{2}H_{\mbox{\tiny bin}}\big(\frac{1+D}{2}\big)+\frac{3}{2}H_{\mbox{\tiny bin}}\big(\frac{3+D}{6}\big)
\label{rhs1.1}
\ee
where $D\equiv \sqrt{1-\Delta_{AA}^2}$. Since many familiar uncertainty relation could be deduced from EUR (\ref{eur2}), such as position-momentum uncertainty \cite{H1,H4}, we believe that our results provide a novel mean to investigate the Hawking effect in a general framework.

\section{Uncertainty game between two static players}
\label{3}

In previous section, we demonstrated that the increment on the entropic uncertainty bound (\ref{rhs1}) due to the presence of a black hole. The Hawking radiation associated with the event horizon just provides an "environment decoherence" on the entanglement between the qubit to be measured and a quantum memory. This phenomenon can be better understand in a framework of open quantum system \cite{YU2}, where a static detector subjected to a bath of fluctuating scalar fields outside the black hole and the Hawking radiation plays the role of vacuum noise. Interestingly, in certain circumstances \cite{YU1,YU5,YU3,YU4}, heat bath can enhance entanglement rather than destroy it, which means initially mutually independent systems can become quantum correlated after long enough time evolution in the bath. 

In this section, we explore an alternative uncertainty game between two static players outside Schwarzschild black hole. The quantum system $A$ to be measured by Alice and the quantum memory $B$ hold by Bob are modeled by two two-level atoms which are in interaction with a bath of fluctuating massless quantum scalar fields outside the black hole. After the total system (atoms and external fields) reaches a equilibrium, the quantum information of $A$ is transferred and stored in quantum memory through entanglement generation between them, which could be witnessed by the negative conditional von Neumann entropy $S(A|B)$ \cite{EUR5} that may significantly reduce the entropic uncertainty bound.

\subsection{Master equation}

Following the discussion in \cite{YU1}, the total system Hamiltonian is
\be
H=\frac{\omega_0}{2}\sum_{i=1}^3n_i\Sigma_i+H_\Phi+H_I
\ee
where $\Sigma_i\equiv\sigma_i^A\otimes\mathbf{1}^B+\mathbf{1}^A\otimes\sigma_i^B$ are symmetrized two-system operators, $\omega_0$ is energy level spacing of the atoms, $\lambda$ is a dimensionless coupling constant. $H_\Phi$ is the Hamiltonian of free massless scalar fields satisfying Klein-Gordon equation outside a black hole, and can be expanded as $\Phi_\mu(x)=\sum_{a=1}^N[\chi_\mu^a\phi^{(-)}(x)+(\chi_\mu^a)^*\phi^{(+)}(x)]$, with $\phi^{(\pm)}(x)$ are positive and negative energy field operators and $\chi_\mu^a$ are the corresponding complex coefficients. $H_I=\lambda \sum_{\mu=0}^3[(\sigma_\mu^A\otimes\mathbf{1}^B)\Phi_\mu(t,\mathbf{ x}_1)+(\mathbf{1}^A\otimes\sigma_\mu^B)\Phi_\mu(t,\mathbf{ x}_2)]$ describes the interaction between the atoms and the bath.

We study the dynamic evolution of the total density matrix $\rho_{tot} = \rho_{AB}(0) \otimes |0\ra\la0|$ in the static frame of atoms, where $ \rho(0)_{AB}$ is the initial reduced density matrix of the two-atom system, and $|0\ra$ is the vacuum state of field $\Phi(x)$. In a weak coupling limit, the two-atom system density matrix $\rho_{AB}(\tau)=\mbox{Tr}_{\Phi}\rho_{tot}(\tau)$ is seen evolving in time according to a one-parameter quantum dynamical semigroup of completely positive maps, generated by the Kossakowski-Lindblad form \cite{YU1,YU3}
\be
\frac{\partial\rho_{AB}(\tau)}{\partial \tau}=-i[H_{\tiny\mbox{eff}},\rho_{AB}(\tau) ]+L[\rho_{AB}(\tau)]
\label{master}
\ee
where $\tau$ is the proper time of static atoms. The effective Hamiltonian is
\be
H_{\tiny\mbox{eff}}=\frac{\omega_0}{2}\sum_{i=1}^3n_i\Sigma_i-\frac{i}{2}\sum_{\alpha,\;\beta=A,B}\sum_{i,j=1}^3H_{ij}\sigma_i^{(\alpha)}\sigma_j^{(\beta)}
\ee
with the coefficients $H_{ij}$ determined by the Hilbert transformation of standard Wightman function $G_{ij}(x-y)=\la0|\Phi_i(x)\Phi_j(y)|0\ra$ which gives $
\mathcal{K}_{ij}(\lambda)=\frac{1}{i\pi}\mbox{P}\int_{-\infty}^{\infty}d\omega\frac{\mathcal{G}_{ij}(\omega)}{\omega-\lambda}
$, a principle integral of Fourier transformation on $G_{ij}$, i.e., $\mathcal{G}_{ij}(\lambda)=\int_{-\infty}^{\infty}d\tau e^{i\lambda\tau}G_{ij}(\tau)$. The Lindbladian operator in (\ref{master}) is given by
\be
L=\sum_{\alpha,\;\beta=A,B}\sum_{i,j=1}^3\frac{C_{ij}}{2}\big[2\sigma_j^{(\beta)}\rho_{AB}\sigma_i^{(\alpha)}-\{\sigma_i^{(\alpha)}\sigma_j^{(\beta)},\rho_{AB}\}\big]
\ee
where the Kossakowski matrix $C_{ij}$ can be written as
\be
C_{ij}=A\delta_{ij}-iB\epsilon_{ijk} n_k+Cn_in_j
\label{op3}
\ee
with
\bea
A&=&\frac{1}{2}[\mathcal{G}(\omega_0)+\mathcal{G}(-\omega_0)],\quad C=\mathcal{G}(0)-A,\no\\
 B&=&\frac{1}{2}[\mathcal{G}(\omega_0)-\mathcal{G}(-\omega_0)]
\eea
 
\subsection{Entanglement generation and entropic uncertainty bound}

We now turn to an uncertainty game between two static observer Alice and Bob, each holding a two-level atom as the qubit to be measured and quantum memory. Since what we concern is the quantum-memory-assisted entropic uncertainty bound of the atom system after its evolution to an equilibrium state in a finite time, we should first demonstrate that how the entanglement could be generated between quantum memory $B$ and the qubit $A$ to be measured. By expressing the reduced density matrix of the two-atom system as
\be
\rho_{AB}(\tau)=\frac{1}{4}[\mathbf{1}^A\otimes\mathbf{1}^B+\sum_{i=1}^3\rho_{i}\Sigma_i+\sum_{i,j=1}^3\rho_{ij}\sigma_i^A\otimes\sigma_j^B]
\label{op2}
\ee
and inserting it back into (\ref{master}), the reduced density matrix at equilibrium is \cite{YU3}
\bea
\rho_{i}&=&-\frac{K}{3+K^2}(\tau+3)n_i,\no\\
\rho_{ij}&=&\frac{1}{3+K^2}[K^2(\tau+3)n_in_j+(\tau-K^2)\delta_{ij}]
\eea
where $K=B/A$ is the ratio of the two constants appearing in the Kossakowski matrix, whose positivity implies $0\leqslant K\leqslant 1$. One can also observe that the final equilibrium state is depend on the choice of initial state by $\tau=\sum_i\rho_{ii}(0)$ which is a constant of motion and satisfies $-3\leqslant\tau\leqslant1$ to keep $\rho_{AB}(0)$ positive. To make the discussion more concise, we come to a simple example with a unit vector as $\vec{n}=(0,0,1)$ which gives the equilibrium state
\be
\rho_{AB}
=\left(\begin{array}{cccc}
\frac{(3+\tau)(K-1)^2}{4(3+K^2)} & 0 & 0 & 0 \\
0 & \frac{3-\tau-(\tau+1)K^2}{4(3+K^2)} & \frac{2(\tau-K^2)}{4(3+K^2)} & 0 \\
0 & \frac{2(\tau-K^2)}{4(3+K^2)} & \frac{3-\tau-(\tau+1)K^2}{4(3+K^2)} & 0 \\
0 & 0 & 0 & \frac{(3+\tau)(K+1)^2}{4(3+K^2)}\end{array}\right)
\label{op1}
\ee
In order to determine whether the final equilibrium state is entangled or not, we calculate the negativity \cite{nega} which is a measure of distillable entanglement and is defined as $N(\rho)=\frac{1}{2}\sum_i(|\lambda_i|-\lambda_i)=-\sum_{\lambda_i<0}\lambda_i$, where $\lambda_i$ are the negative eigenvalues of partial transposed density matrix of (\ref{op1}). The value of negativity ranges from 0, for separable states, to 0.5, for maximally entangled states. Form  (\ref{op1}),  we can straightforwardly obtain
\be
N=\frac{2\sqrt{K^4+\tau^2+K^2(9+4\tau+\tau^2)}-(3+\tau)(1+K^2)}{4(3+K^2)}
\label{negativity}
\ee 
which reaches maximum 0.5 at $\tau=-3$ and is vanishing at $\tau=\frac{5K^2-3}{3-K^2}$, agreeing with \cite{YU3}. 

To calculate the entropic uncertainty bound in r.h.s. of (\ref{eur2}), we give the von Neumann entropy of (\ref{op1}) which is $S(\rho_{AB})=-\sum_{i=1}^4\lambda_i\log_2\lambda_i$ with eigenvalues
\bea
\lambda_1&=&\frac{(1+K)^2(3+\tau)}{4(3+K^2)},\quad\lambda_2=\frac{(1-K)^2(3+\tau)}{4(3+K^2)}\no\\
\lambda_3&=&\frac{(1-K^2)(3+\tau)}{4(3+K^2)},\quad\lambda_4=\frac{1-\tau}{4}
\eea
From the definition (\ref{op2}), the quantum memory $B$ hold by Bob is given by $\rho_B=\mbox{Tr}_A\rho_{AB}=\frac{1}{2}\big(\mathbf{1}^B-\frac{K(3+\tau)}{3+K^2}\sigma_3^B\big)
$, 
corresponding to a binary entropy $S(\rho_B)=H_{\tiny\mbox{bin}}\big(\frac{K^2-(3+\tau)K+3}{2(3+K^2)}\big)$. Finally, we can get the entropic uncertainty bound from (\ref{eur2}) for the uncertainty game implemented when the two-atom system evolves to an equilibrium state, which is 
\bea
U_b&=&1+\frac{3+\tau}{4}\log_2(3+K^2)+H_{\tiny\mbox{bin}}\bigg(\frac{1-\tau}{4}\bigg)-H_{\tiny\mbox{bin}}\bigg(\frac{K^2-(3+\tau)K+3}{2(3+K^2)}\bigg)\no\\
&&-(3+\tau)\sum_{i=0,1}\bigg[\frac{1}{4}+\frac{(-1)^iK}{3+K^2}\bigg]\log_2[1+(-1)^iK]
\label{rhs2}
\eea

To demonstrate the possible violation of uncertainty bound $-\log_2c$, it is particularly important to choose a proper initial state $\rho_{AB}(0)$. In our case, we need to select a initial state that the unavoidable Hawking decoherence would not be sufficient to counteract entanglement generation after the dynamical evolution of system, thus leaves an equilibrium state gives $U_b<1$. For example, we can start with a Bell-diagonal state with $c_1=c_2=c_3=-0.75$ corresponding to $\tau=-2.25$. Implementing the uncertainty game with this initial state, one obtains the intrinsic uncertainty bound $U_b=1$ for measurement $\sigma_j$ and $\sigma_k$ with $j\neq k$. In passing from this initial state to its corresponding equilibrium state, the corresponding increase in negativity can be easily computed from (\ref{negativity}) as 
\be
\delta N=N[\rho]-N[\rho(0)]=\frac{\sqrt{(4K^2+9)^2+9K^2}-(4K^2+9)}{8(3+K^2)}
\label{nincre}
\ee
and the uncertainty bound (\ref{rhs2}) becomes
\bea
U_b&=&1+H_{\tiny\mbox{bin}}\bigg(\frac{13}{16}\bigg)-H_{\tiny\mbox{bin}}\bigg(\frac{4K^2-3K+12}{8(3+K^2)}\bigg)+\frac{3}{16}\log_2(3+K^2)\no\\
&&-\sum_{i=0,1}\bigg[\frac{3}{16}+\frac{(-1)^i3K}{4(3+K^2)}\bigg]\log_2[1+(-1)^iK]
\label{rhs3}
\eea
If Kossakowski coefficients lead to $U_b<1$, we conclude that the entanglement generation during the dynamical evolution of  the total system (atoms and external fields) can efficiently lower the uncertainty bound.

\subsection{Boulware vacuum and Hartle-Hawking vacuum}

In order to explore the behavior of above uncertainty bound in a specific curved background, we need to calculate the Kossakowski coefficients (\ref{op3}), which determined by the scalar field correlation function in the vacuum state. For Schwarzschild spacetime, as mentioned in previous section, one can at least define two vacuum states, i.e., Boulware and Hartle-Hawking vacuum states. This requires us examine uncertainty bound (\ref{rhs2}) in these two vacua, respectively.

\begin{figure}[hbtp]
\begin{center}
\includegraphics[width=.45\textwidth]{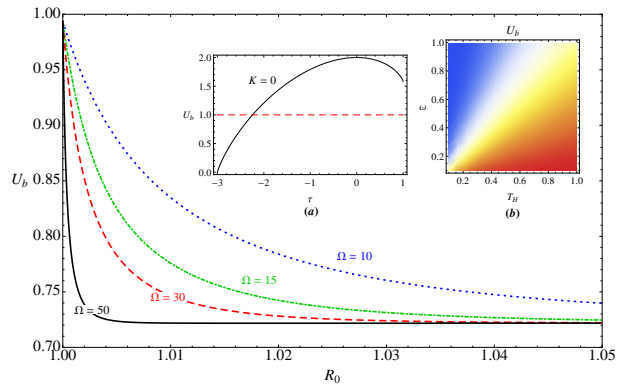}
\caption{The uncertainty bound $U_b$ is a function of $\omega$, Hawking temperature $T_H$ and $R_0$. After quantum dynamical evolution, two static two-level atoms (modeling the qubit to be measured and a quantum memory), can be more entangled, therefore reduces the entropic uncertainty bound for incompatible measurements. In the inset (a), the dependence of $U_b$ on the choice of initial state is demonstrated for $K=0$. In the inset (b), we plot the asymptotic behavior of uncertainty bound at $R_0\rightarrow\infty$, where the red color denotes higher $U_b$. }
\label{ubh3}
\end{center}
\end{figure}

For Hartle-Hawking vacuum, the Wightman function is \cite{WITM,DEWITT}
\be
G^+_H(x,x')=\sum_{lm}\int^\infty_0\frac{|Y_{lm}(\theta,\phi)|^2}{4\pi\omega}\bigg[\frac{e^{-i\omega\Delta t}}{1-e^{-2\pi\omega/\kappa}}|\overrightarrow{R}_l(\omega,r)|^2+\frac{e^{i\omega\Delta t}}{e^{2\pi\omega/\kappa}-1}|\overleftarrow{R}_l(\omega,r)|^2\bigg]d\omega
\ee
whose Fourier transformation is 
\bea
\mathcal{G}(\lambda)&=&\int_{-\infty}^\infty e^{i\lambda \tau} G^+_H(x,x')d\tau\no\\
&=&\sum_{l}\frac{2l+1}{8\pi\lambda}\bigg[\frac{|\overrightarrow{R}_l(\lambda\sqrt{g_{00}},r)|^2}{1-e^{-2\pi\lambda\sqrt{g_{00}}/\kappa}}+\frac{|\overleftarrow{R}_l(\lambda\sqrt{g_{00}},r)|^2}{1-e^{-2\pi\lambda\sqrt{g_{00}}/\kappa}}\bigg]\no
\eea
Then the Kossakowski coefficients can be given explicitly close to the event horizon and at infinity \cite{YU1}. By using the geometrical optics approximation \cite{DEWITT}, we finally have a simple result
\be
K=\frac{e^{2\pi\omega_0\sqrt{g_{00}}/\kappa}-1}{e^{2\pi\omega_0\sqrt{g_{00}}/\kappa}+1}=\tanh\bigg[\Omega\sqrt{1-\frac{1}{R_0}}\bigg]
\label{Rval}
\ee
Here we introduce the same parameterization as in (\ref{bh3}). Plugging it into (\ref{rhs3}), we have the entropic uncertainty bound for two static atoms in a bath of fluctuating scalar field in the Hartle-Hawking vacuum, which is depicted in Fig.\ref{ubh3}.

At close to the horizon, i.e., when $r_0\rightarrow R_H$, $K \rightarrow 0$, we obtain $\delta N=0$ which means that no entanglement between the final states of the quantum memory and system to be measured could be generated. From (\ref{rhs3}) we obtain $U_b=1$ indicating an intrinsic uncertainty. 

However, when $r_0$ grows, we have a nonzero entanglement increment $\delta N>0$. As mentioned before, this suggests the quantum information of qubit $A$ has been transferred and stored in quantum memory after the dynamical evolution of total system, therefore induces the uncertainty bound. From Fig. \ref{ubh3}, we note that for atoms with fixed energy spacing, higher Hawking temperature $T_H$ introduces additional uncertainty in $U_b$ as Hawking decoherence counteract part of entanglement generation. Conversely, for a Schwarzschild black hole with fixed mass, observer can choose his atom with large energy spacing to get more precisely prediction on measurement outcomes. As both observers leave away from the horizon, the influence of Hawking effect on the measurement uncertainty could be ignored. In the asymptotic case $r_0\rightarrow \infty$, (\ref{Rval}) approaches a maximal value $K\rightarrow\tanh(\Omega)$, the dependence of uncertainty bound (\ref{rhs3}) on Hawking temperature is depicted in Fig. \ref{ubh3} (b).

For Boulware vacuum, the Wightman function of scalar field is \cite{WITM}
\bea
G^+_B(x,x')=\sum_{lm}\int^\infty_0\frac{e^{-i\omega\Delta t}}{4\pi\omega}|Y_{lm}(\theta,\phi)|^2[|\overrightarrow{R}_l(\omega,r)|^2+|\overleftarrow{R}_l(\omega,r)|^2]d\omega
\eea
The corresponding Fourier transformation with respect to proper time reads 
\be
\mathcal{G}(\lambda)
=\sum_{l}\frac{2l+1}{8\pi\lambda}[|\overrightarrow{R}_l(\lambda\sqrt{g_{00}},r)|^2+|\overleftarrow{R}_l(\lambda\sqrt{g_{00}},r)|^2]\theta(\lambda)\no
\ee
where $\theta(\lambda)$ is the step function. Then the Kossakowski coefficients are \cite{YU1}
\be
A=B=\sum_{l=0}^\infty\frac{2l+1}{16\pi\omega}[|\overrightarrow{R}_l(\omega,r)|^2+|\overleftarrow{R}_l(\omega,r)|^2]
\ee
which gives $K=1$. We can easily obtain the entanglement generation in this case as $\delta N\sim0.011$, which indicates that after the total system reaches a equilibrium with respect to Boulware vacuum, the quantum information of qubit $A$ has been transferred and stored in quantum memory. Interestingly, in this case, we find that the entropic uncertainty bound can be reduced to $U_b\sim0.722$ everywhere outside the event horizon. This is just because there is no thermal radiation present with respect to the Boulware vacuum, which indicates the same amount of entanglement generation as that in Minkowski vacuum \cite{YU5}.

\section{Summary and discussion}
\label{4}

In this Letter, we investigate quantum-memory-assisted EUR in the curved background outside a Schwarzschild black hole. We find a nontrivial relation between uncertainty bound (\ref{rhs1}) and Hawking effect in two sample uncertainty games. For an uncertainty game between a free falling observer and her static partner who holds a quantum memory initially entangled with the qubit to be measured, we show that in the view of static observer, the entropic uncertainty bound (\ref{rhs1}) increases due to entanglement degradation provoked by Hawking radiation. To save this, the static observer should be placed at a far distance from the event horizon or choose a quantum memory with high enough mode frequency. On the other hand, since entanglement between static systems could be generated through a bath of fluctuating external field, by modeling the quantum particle to be measured and the quantum memory with two static atoms interacting with an external scalar field, we find Hawking decoherence can be overcome and leads to a lower entropic uncertainty bound (\ref{rhs3}) after the quantum dynamical evolution of atoms system. We show that the uncertainty bounds for both games are dependent on three physical parameters, i.e., the mass of the black hole, the distance of observer from event horizon, and the mode frequency of quantum memory. Moreover, for open quantum system, uncertainty bound is also sensitive with the choice of initial bipartite state of atoms. Numerically estimation for a proper choice of initial state shows that the reduction of uncertainty bound is comparable with possible real experiments \cite{EUR4}. 

Our study raises several implications. Firstly, since Hawking radiation always plays a role as an environment decoherence, the affected uncertainty bound should certainly be a result of competition between Hawking decoherence and the entanglement evolution of uncertainty game system. Once generalize to a cosmological background \cite{FENG1}, additional entanglement generation from the dynamical evolution of spacetime should be taken into account \cite{COS}. Since uncertainty bound is a sort of measure of quantumness \cite{OPP}, it is reasonable to expect that such investigation of (\ref{eur2}) in a cosmological frame may shed new light on the quantum-to-classical transition in early universe \cite{POL}. 

Secondly, to illustrate the generality of our results, we employed AA time-energy uncertainty (\ref{AA1}) as building block to reexpress the uncertainty bound. Although we demonstrated this relation in a specific case, a rigorous link between two uncertainty measure could be established. For instance, in a geometrical view, AA time-energy uncertainty is a statement about the velocity of the Killing flow on $\mathbb{C}\mathbf{ P}^n$ manifold with Fubini-Study metric \cite{GEO1}, which defines a distance measure on manifold satisfies a triangle inequality. For pure state, this gives geometric derivation of entropic uncertainty relation \cite{GEO2}. The generalization of this approach might prove a rigorous link between entropic uncertainty and other uncertainty probe.

Finally, while the methodology we adopted in this Letter is compatible with common interpretation on Hawking decoherence \cite{RQI1,FIREWALL1}, Almheiri \emph{et al.} (AMPS) recently revealed \cite{FIREWALL2} a profound conflict between core quantum properties of black holes and the equivalence principle, indicates a sharp paradox that an observer falling into a black hole would either witness a violation of Entanglement Monogamy (EM) or burn up by a firewall structure erecting around the event horizon. Despite the rapid and intensive response of physics community (see \cite{FIREWALL3} for instance), ranging from skepticism to ambivalence to endorsement, this firewall paradox is still an unsettled research question. While an overall struggle with the issue is beyond the scope of present Letter, here we make a first move by relating the paradox to an entropic uncertainty relation. To proceed, recall that the stretching spacetime around the horizon creates correlated Hawking pair $\{b_i,c_i\}$ at each time step $t_i$. AMPS argued \cite{FIREWALL2} that after the black hole evaporate more than half mass, the resulting early Hawking radiation $A\equiv \{b_1,\cdots,b_n\}$ should be entangled with a late radiated quanta $B\equiv b_{n+1}$ by unitarity \cite{Page} and also with inside quanta $C\equiv c_{n+1}$ by equivalence principle, thus violates monogamy of quantum entanglement. In an entropic view, AMPS just adapted that the entanglement entropy of the emitted quanta necessarily increases with each emission, i.e., $S(B|A)\geqslant\ln2$, which contradicts the traditional complementarity that entropy must decrease after the halfway point via the unitarity postulate \cite{FIREWALL4}. One should note that the AMPS argument indeed implies an observer falling into a black hole can perform a feasible experiment to witness the violation of EM. For tripartite state $\rho_{ABC}$, taking advantage of the monogamy of entanglement, EUR (\ref{eur2}) can be generalized to a tripartite uncertainty relation for black hole as \cite{EUR1}
\be
S(Q|B)+S(R|C)\geqslant-\log_2c
\label{eur3}
\ee
where $B$ and $C$ can be viewed as two separate quantum memories. The conditional von Neumann entropy $S(X|B)$ quantifies the uncertainty after the measurements $X\in(Q,R)$ are performed. The post measurement state is $\rho_{XB}=\sum_x(\Pi_x\otimes\mathbf{1})\rho_{AB}(\Pi_x\otimes\mathbf{1})$, where $\Pi_x=|\psi_x\ra\la\psi_x|$ and $\{|\psi_x\ra\}$ are the eigenstates of the observable $X$. This inequality can be interpreted for a black hole as follows. If one can use a late radiated quanta to predict the outcome of the $Q$ measurement with certainty (i.e. $S(Q|B) = 0$), then the necessarily uncertainty about the outcome of the $R$ measurement assisted by inside quanta  (i.e. $S(R|C) > 0$) is inevitable as long as the measurements are incompatible (i.e. $c < 1$). The AMPS argument can now be translated into the context of uncertainty principle, and suggests that the observer falling into a black hole would either hit the firewall or perform a proper uncertainty game to witness the violation of the tripartite uncertainty relation (\ref{eur3}). In this meaning, further exploration on a subtle uncertainty game involving the firewall paradox may shed new light on our understanding of the field. We will report the related work elsewhere.

\section*{Acknowledgement}
This work is supported by the ARC through DP140101492. We thank the anonymous referee for the insightful suggestion to make contact with the firewall paradox. J. F. thanks Bo Li and Zhao-Long Wang for stimulating discussions. H. F. acknowledges the support of NSFC 973 program through 2010CB922904.

\end{document}